\documentclass[aps,prl,twocolumn,superscriptaddress]{revtex4}

\usepackage{graphicx}
\usepackage{amssymb}
\usepackage{amsmath}
\usepackage{epstopdf}
\usepackage{mciteplus}
\usepackage{setspace}

\begin{document}


\title{Structural anisotropy of directionally dried colloids}

\author{Fran\c{c}ois Boulogne}
\author{Ludovic Pauchard}
\author{Fr\'ed\'erique Giorgiutti-Dauphin\'e}
\affiliation{FAST, UPMC, Univ Paris-Sud, CNRS UMR 7608, F-91405, Orsay, France}
\author{Robert Botet}
\affiliation{Physique des Solides, CNRS UMR 8502, Univ Paris-Sud, F-91405 Orsay, France}
\author{Ralf Schweins}
\affiliation{DS/LSS Group, Institut Laue-Langevin, BP156, F-38042 Grenoble Cedex 9, France}
\author{Michael Sztucki}
\affiliation{ESRF, B.P. 220, 38043 Grenoble, France}
\author{Joaquim Li}
\author{Bernard Cabane}
\affiliation{PMMH, CNRS UMR 7636, ESPCI, 10 rue Vauquelin, 75231 Paris Cedex 05, France}

\author{Lucas Goehring}
\email[]{lucas.goehring@ds.mpg.de}
\affiliation{Max Planck Institute for Dynamics and Self-Organization (MPIDS), 37077 G\"ottingen, Germany}


\begin{abstract}
Aqueous colloidal dispersions of silica particles become anisotropic when they are dried through evaporation. This anisotropy is generated by a uniaxial strain of the liquid dispersions as they are compressed by the flow of water toward a solidification front. Part of the strain produced by the compression is relaxed, and part of it is stored and transferred to the solid. This stored elastic strain has consequences for the properties of the solid, where it may facilitate the growth of shear bands, and generate birefringence.
\end{abstract} 


\maketitle

\begin{figure}
\includegraphics[width=3.45in]{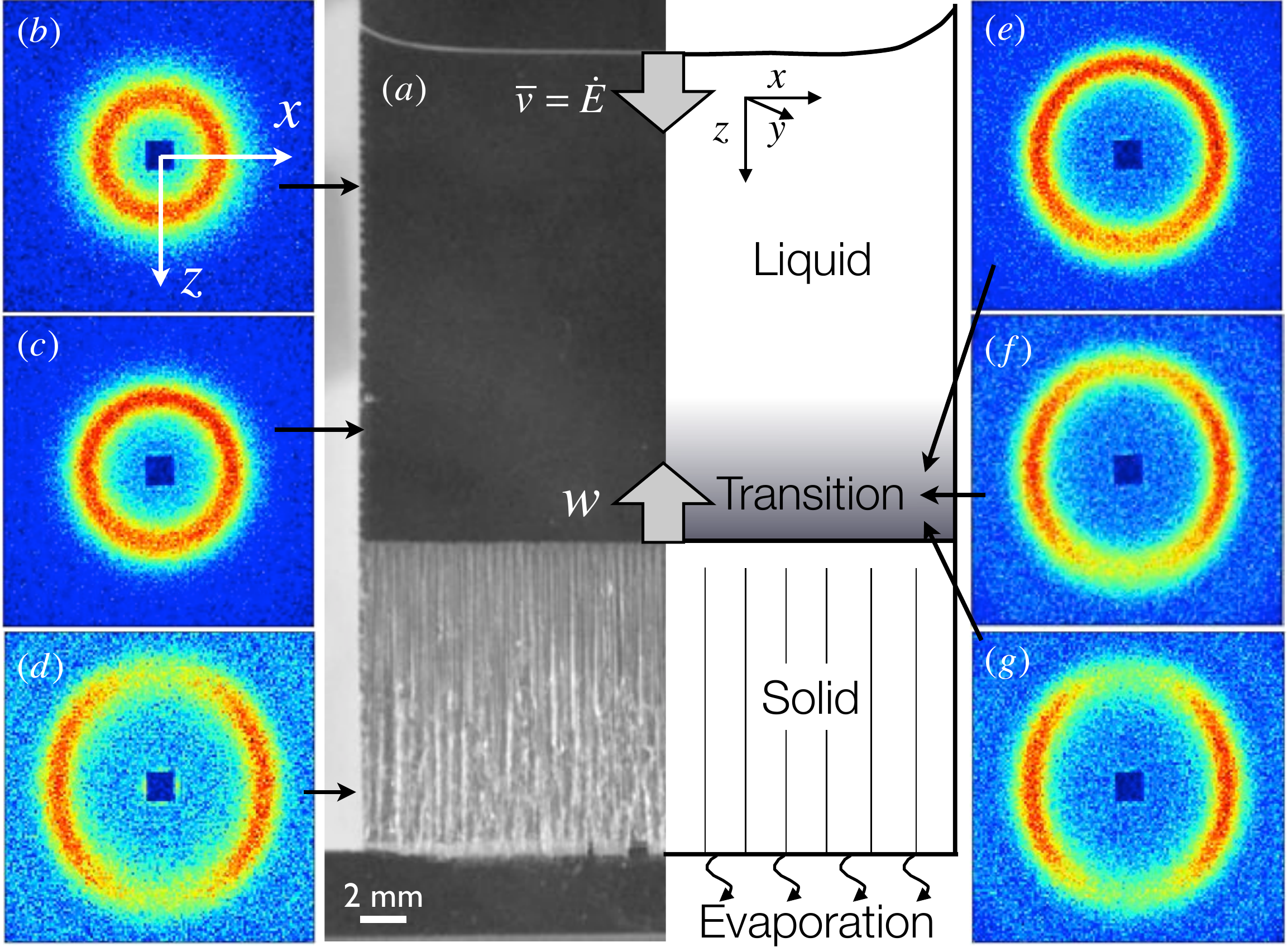}
\caption{\label{fig1} Colloidal dispersions were dried in Hele-Shaw cells, with evaporation proceeding from one open end, and spacers on either side.  (a) The upper meniscus of the dispersion receded to balance evaporation, while the solid phase was left in a rigid deposit that grew upward at velocity $w$.  In the liquid region far from the solidification front, the scattering profile (b) is the ring of a colloidal liquid, which (c) increases in radius as this front is approached.  In the solid region (d), the ring is stretched in the $z$ direction, along which the structure is compressed by $\sim10\%$.  (e-g) This structural anisotropy arises within a transition region prior to solidification.}
\end{figure}

\section{Introduction}
There are two main routes for making materials. For homogeneous materials such as metals the main route is through the liquid-solid transition that results from cooling, as intermolecular forces cause atoms or molecules to settle into equilibrium positions. For particulate materials such as ceramics and coatings, the usual route is through the liquid-solid transition that is caused by evaporation from a dispersion of solid particles in a volatile solvent \cite{Chiu1993a,Allain1995,Dufresne2003,Routh2013}. One often sees this transition as being driven exclusively by the loss of free volume \cite{Pusey1986}, and thus the solvent disappears from the description. However, there are many cases where the transport of solvent during solidification plays a role that is important and counterintuitive.  For example in the transport of particles to the edge of a drying drop \cite{Deegan1997}, or in the formation of drying fronts in colloidal films \cite{Dufresne2003,Goehring2010,Routh2013}.  

Here we show that the flow of solvent in a directionally dried colloidal dispersion breaks the orientational symmetry of the liquid and generates a structural anisotropy. This anisotropy arises when the particles have been concentrated by the flow to the point where they are caged by their neighbours into a soft, deformable network. It is a robust feature of directional drying, which we have observed for different particle sizes and salt concentrations. The anisotropy remains as a residual strain in the solid, where it causes birefringence, and may drive shear-banding. 

\section{Methods} 
\begin{figure}
\includegraphics[width=3.45in]{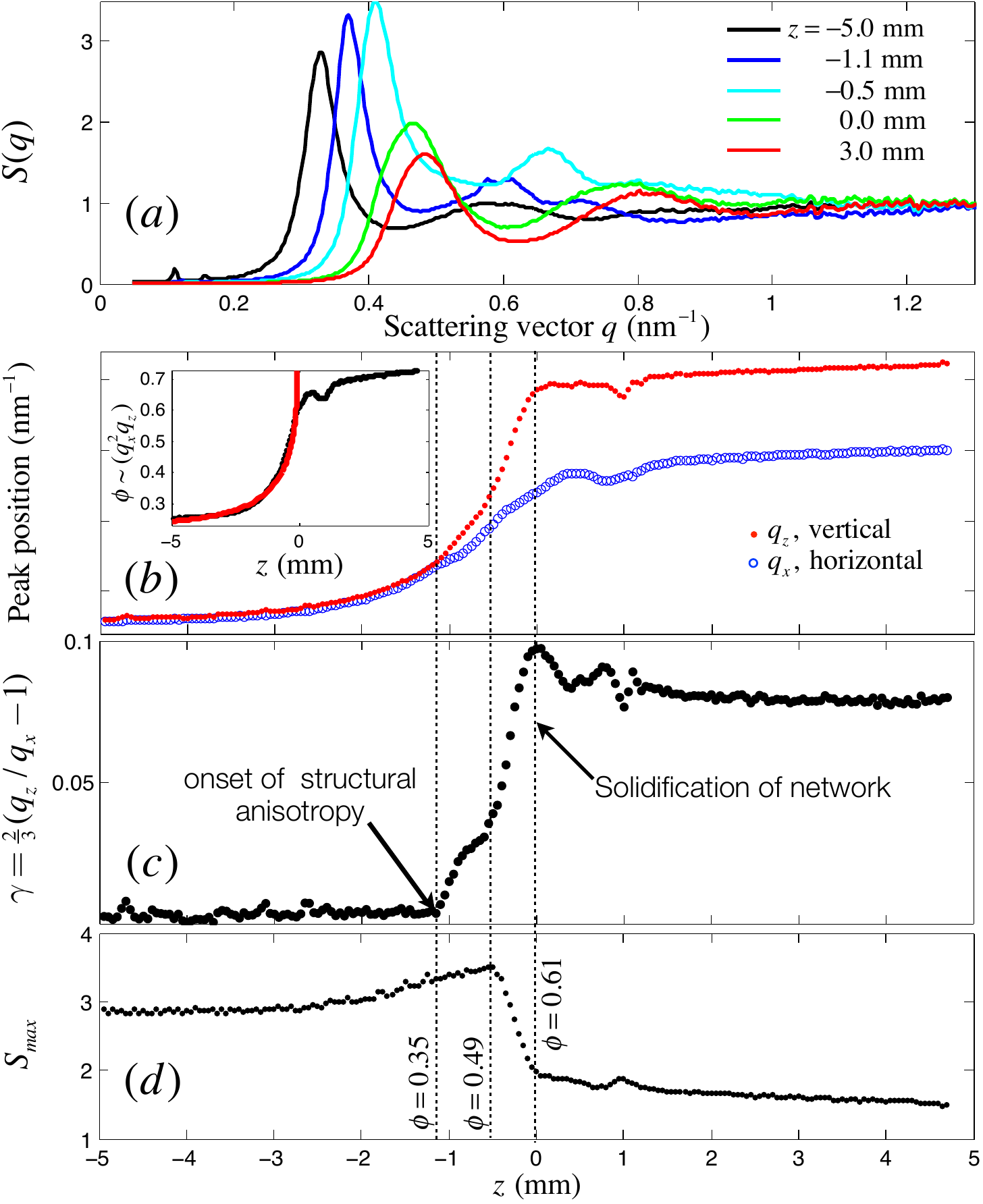}
\caption{\label{fig2}  A drying dispersion of 8 nm particles in 5 mM NaCl, seen by SAXS.  (a) The azimuthally averaged structure factor $S(q)$ evolves as the sample is scanned across the solidification front ($z=0$). (b) The position of the structure factor peak increases as the dispersion concentrates.  The corresponding increase in volume fraction (inset, black points) is well-described by the balance between advection along $z$, and diffusion back down concentration gradients (red curve, Eqn. \ref{diff2}). 1.1 mm prior to solidification the peak positions along the $x$ (open blue circles) and $z$ (red points) axes diverge.   The scattering is anisotropic past this point, and (c) the deviatoric strain $\gamma$ increases until solidification.  In the solid, $\gamma$ drops slightly, as cracks open parallel to the $z$-axis. (d) The height $S_{max}$ of the peak of $S(q)$ increases slowly through the transition region, until 0.5 mm before solidification, when it collapses, suggesting the onset of partial aggregation. Solidification, $z=0$, is inferred from the stabilisation of $S_{max}$, $q_x$ and $q_z$, which coincide with the maximum $\gamma$. }
\end{figure}

Our experiments consist of small-angle neutron (SANS) and x-ray (SAXS) scattering through aqueous dispersions of charged colloidal silica drying in Hele-Shaw cells, as in Fig. \ref{fig1}(a). We used Levasil-30 (Azko-Nobel, particle radius $a = 46$ nm) as received, and dialysed Ludox SM-30 ($a = 5$ nm), HS-40 ($a = 8$ nm), and TM-50 ($a = 13$ nm) against aqueous solutions of NaCl and NaOH (pH 9.5), as in Ref. \cite{Li2012}.  Poly(ethylene glycol) was added to the dialysis buffer, to concentrate the dispersions by osmotic stress.   For SANS, cells were made of two 100$\times$50 mm quartz glass plates separated by 0.3 mm plastic spacers along two edges, with evaporation occurring along the open bottom of the cell.  For SAXS the quartz was replaced by mica, held rigid by glass supports around narrower (3-5 mm) openings. All cells were filled with dispersions of initial volume fraction $\phi_0$ from the bottom, and laid near-horizontally. Evaporation, at rate $\dot E$ (volumetric rate per unit area), from the open end of the cell was balanced by a retreat of the upper meniscus within the cell, at velocity $\bar{v} = \dot E$.  For instance, for the experiment highlighted in Fig. \ref{fig2}, $\bar{v}=0.4$ $\mu$m/s. A band of wet aggregated solid, of final volume fraction $\phi_f$ appeared near the cell's lower end, and grew upwards at velocity $w = -\bar{v}\phi_0/(\phi_f-\phi_0)$, as in Fig. \ref{fig1}. This band could be seen by a change in opacity corresponding to the irreversible solidification of particles into a rigid connected network \cite{Goehring2010}.   After a few hours of drying, cells were raised vertically, and set in the path of a neutron or x-ray beam.

We used small-angle scattering to scan the structures of dispersions along lines normal to the solidification front. SAXS was performed with ID02 at ESRF, SANS with D11 at ILL. SAXS spectra were collected every 2.5 s, with a step size and beam $z$-resolution of 50 $\mu$m. SANS spectra were collected every 400 s, with a $z$-resolution of 600 $\mu$m. While the motion of the drying fronts was negligible (1-2\% of step size) during SAXS scans, they were noticeably (10-20\%) foreshortened during SANS scans.  Only SAXS data are used to analyse the front shape. 

The 2D scattering patterns show a high-intensity ring that reflects the structure of the shell of nearest neighbours around each particle in the irradiated volume. Some SANS patterns are shown in Fig. \ref{fig1}(b-g). Since the scattering was anisotropic, we regrouped these patterns over scattering vectors $q$ within $\pm5^\circ$ of the $x$ and $z$ axes, or azimuthally. By rotating the cell around the $z$ axis (flow direction) we found that both perpendicular directions, $x$ and $y$, were equivalent, $q_x\simeq q_y$. We calculated structure factors by dividing the regrouped scattering intensities with the form factor of a corresponding dilute dispersion. The position of the main structure factor peak in the flow direction, $q_z$, and perpendicular to the flow, $q_x$, show how the structure changes (Figs. \ref{fig1}, \ref{fig2}). We used these to estimate volume fractions of the particles in the colloidal dispersion as $\phi = \alpha(q_x^2q_z)$. For each dispersion, $\alpha$ is fit to spectra from a series of sealed calibration samples of known $\phi$; for the 8 nm silica, $\alpha= 7.04$ nm$^3$ based on samples from $\phi= 0.05$ to 0.49 \cite{Li2012}.  We extrapolate beyond this calibration range, although this extrapolation does assume that the average shape of the unit cell does not significantly change, other than through compression (along any axes).  

\section{Observations and Discussion} Scanning across the liquid-solid transition of the drying cell, we found three distinct regions in which the structural properties of the colloidal dispersion were qualitatively different: a colloidal liquid ($\phi<0.35$), a soft colloidal glass ($0.35\leq\phi<0.61$), and a wet aggregated solid ($\phi\geq 0.61$).   The colloidal glass transition is determined by the sudden appearance of a finite shear modulus, and occurs at lower volume fractions than in hard-sphere colloidal dispersions \cite{Pusey1986}, due to the small size and large charge of our dispersions.  A distinct feature of the intermediate colloidal glass is that its structure remains anisotropic after uniaxial compression, as we will soon show.  

The uppermost part of the cell had structural features that were characteristic of a colloidal liquid.  The changes observed here will serve as a benchmark to characterise the behaviour in the two later regions.  The scattering pattern was a circular ring, whose expansion reflects an increasing concentration of the particles as they approach the liquid-solid transition. From the peak positions (Fig. 2) we calculated that $\phi$ increased from 0.23 ($\phi_0$, initial volume fraction) to 0.35, for 8 nm particles in 5 mM NaCl, while remaining isotropic. Bulk dispersions of such silica, concentrated through osmotic stress to comparable $\phi$ had the same structure factors \cite{Li2012} and fluid-like properties \cite{Persello1994,Giuseppe2012} as we observed.  The compression of the liquid dispersion can be understood by a simple force balance, as follows.

Prior to the liquid-solid transition, advection towards the solid competes with diffusion back down concentration gradients. The particles slow down, and experience a force from the water that continues to flow past them, towards the evaporating edge of the cell. The drag $F_{d}$ of a fluid past particles of number density $n\sim\phi$ balances gradients in osmotic pressure, $\nabla \Pi = nF_{d}$. $F_d$ is the product of a Stokes drag, and a hindered settling coefficient $r(\phi)$ that captures the hydrodynamical interactions between particles \cite{Buscall1987}.  Other colloidal interactions are included through the compressibility factor $Z(\phi) = \Pi(\phi)/nkT$, where $kT$ is the Boltzmann thermal energy \cite{Peppin2006}.  This model predicts that $\phi$ obeys a nonlinear advection-diffusion equation
\begin{equation}
\label{diff}
{\partial_t\phi} + \nabla(\phi \bar{v}) = \nabla(D(\phi)\nabla\phi)),
\end{equation}
where 
\begin{equation}
D = \frac{D_0}{r(\phi)}\frac{\partial \phi Z}{\partial \phi}
\end{equation}
and $D_0$ is the Stokes-Einstein diffusivity \cite{Peppin2006}. Similar behaviour is seen for dispersed particles in sedimentation \cite{Buscall1987}, freezing \cite{Peppin2006} and filtration \cite{Aimar2010}.  For a dispersion with a far-field velocity $\bar{v}$ approaching a flat liquid-solid front advancing at velocity $w$, in the co-moving reference frame of that front, Eqn. \ref{diff} predicts a steady-state concentration profile that follows
\begin{equation}
\label{diff2}
\phi-\phi_0 = \frac{L}{r}\frac{\partial(\phi Z)}{\partial \phi}\frac{\partial\phi}{\partial z}
\end{equation}
where $L = D_0/(\bar{v}-w)$ is the advection-diffusion length for non-interacting particles.  For 8 nm spheres, taking a measured $(\bar{v}-w)$ = 0.64$\pm$0.1 $\mu$m/s, $L$ = 41$\pm7$ $\mu$m.  Although $L$ is about 20 times smaller than the observed width of the compaction region shown in Fig. 2(b), we find that the larger interaction energies of charged colloids can account for this difference. We calculated the compressibility $Z$ by the Poisson-Boltzmann cell method described in J\"onsson {\it et al.} \cite{Jonsson2011}, where it is shown to be a good model for aqueous silica dispersions. Further taking $r = (1-\phi)^{-6}$ as suggested in \cite{Russel1989,Peppin2006}, we solved Eqn. \ref{diff2} numerically for a far-field $\phi_0 = 0.23$, allowing the solution to diverge at $z=0$.  As shown in Fig. \ref{fig2}(b), prior to solidification $\phi(z)$ is well-fitted by this model of the compression of a charged colloidal fluid, with no free parameters.

Closer ($\sim$1 mm) above the liquid-solid transition, however, we found that the diffraction ring started expanding faster along the flow direction than along the perpendicular direction, as shown in Fig. \ref{fig1}. As a result, it took the shape of an ellipse, corresponding to stronger compression of particles in the flow direction and weaker compression in the perpendicular directions. The structure factor peak positions $q_x$ and $q_z$ show the same behaviour, diverging from each other (Fig. \ref{fig2}(b)) while the dispersion was smoothly increasing in concentration.

\begin{figure}
\includegraphics[width=3.45in]{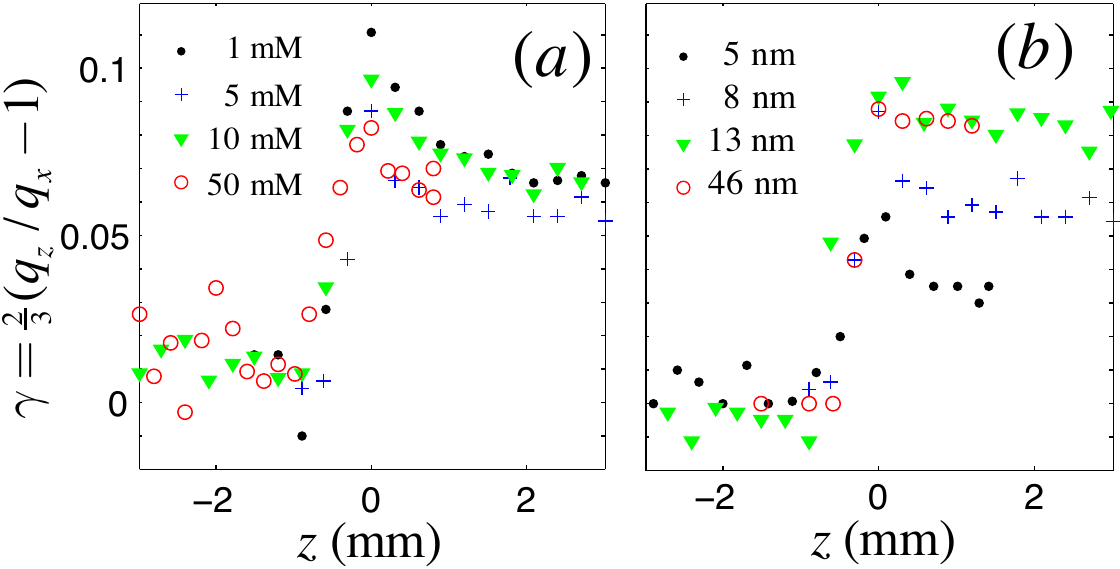}
\caption{\label{fig3} SANS experiments show that (a) changes to the concentration of electrolyte in solution (for 8 nm particles) does not strongly affect the growth of $\gamma$, while (b) particle radius can slightly, in particular for very small particles.}
\end{figure}

An anisotropic structure is not the equilibrium structure of a normal colloidal liquid, which should be able to relax any strains other than hydrostatic compression. It requires a finite shear modulus, a rigidity, of the particle arrangement.  On the other hand, the volume fraction $\phi = \alpha(q_x^2q_z)$ of these dispersions continues to evolve in accordance with a balance of drag forces against the increase in osmotic pressure of the dispersion, as shown in the inset to Fig. \ref{fig2}(b).  Thus, past the point when the anisotropy begins, $\phi_c$, the particle arrangement has a hydrostatic strain $\epsilon_0$ in equilibrium with the osmotic pressure, and an additional deviatoric (shape-changing, but volume preserving) strain.  Since we confirmed, by rotation of the sample, that scattering patterns along the $x$ and $y$ axes are the same, and we are applying a uniaxial compression along the $z$-direction, the appropriate strain tensor is
\begin{equation}
\epsilon_{ij} =  \epsilon_0\delta_{ij} +  \begin{pmatrix} \gamma/2 & 0 & 0 \\
0 &  \gamma/2 & 0 \\
0 & 0 &  -\gamma \end{pmatrix}
\end{equation}
Here, $\gamma$ is the magnitude of the deviatoric strain. The structure factor describes the average arrangement of the shell of neighbours around a particle, while $\gamma$ represents the deformation of this shell.  If the reference state $\gamma=0$ is taken to be that of an isotropic distribution of neighbours, then to leading order, in terms of our observables
\begin{equation}
\gamma = \frac{2}{3}\bigg(\frac{q_z}{q_x}-1\bigg).
\end{equation}
This can be derived by considering the volume-preserving deformation of a sphere into an oblate ellipsoid of major axis $q_x^{-1}$ and minor axis $q_z^{-1}$. The evolution of $\gamma$ for the drying of 8 nm silica in 5 mM NaCl is shown in Fig. \ref{fig2}(c).  It starts suddenly at $\phi_c=0.35$, rises rapidly to a maximum of $\gamma=0.10$ at the point of solidification, and remains permanently high in the aggregated solid.

We observed a similar level of anisotropy ($\gamma\sim 0.1$) during the solidification of every sample we investigated, for drying dispersions made with ionic strengths of 1-50 mM (Fig. \ref{fig3}(a)) and with particles of radii from 5 to 46 nm  (Fig. \ref{fig3}(b)).  The phenomenon appears very robust, always taking place in a range of volume fractions before the liquid-solid transition, and always producing a maximum strain of about 10\% immediately before this transition. 

In order to verify that our dispersions could have the equilibrium $\phi$ but not the isotropic equilibrium structure, we checked the behaviour of dispersions that had been concentrated by osmotic stress. For 8 nm silica in 5 mM NaCl, concentration past $\phi$ = 0.3 produced a crossover in mechanical behaviour from a liquid to a transparent material that retained its shape against the effects of gravity.   Similarly, rheological measurements on undialyzed HS-40 show that it has a yield stress above $\phi$ = 0.32 \cite{Giuseppe2012}, while 10 nm silica in 5 mM NaNO$_3$, dialysed by osmotic stress has shown a yield stress above $\phi=0.24$ \cite{Persello1994}. The persistence of deviatoric strains in the compressed dispersions indicates that they have crossed into a state where the particles are permanently caged by their interactions, and can only relieve stress when the strain on the cage is sufficient to allow a change of neighbours. This is the equivalent of an ergodic to nonergodic glass transition \cite{Mason1996}, and is expected when the pair potential of adjacent particles reaches $\sim kT$ \cite{Russel1989}.  Indeed, the pair potential of 8 nm silica particles in 5 mM NaCl at $\phi$ = 0.35 is about 2 kT at the average inter-particle distance (as calculated through DLVO theory, taking into account both the counterions of the surface sites and the NaCl added through dialysis \cite{Trulsson2009}).  At $\phi$ = 0.61, immediately before the liquid-solid transition, this pair potential reaches 21 kT. Hence, in this intermediate region the dispersion behaves as a soft colloidal glass, as the particle are effectively trapped by their nearest neighbour shells.

\begin{figure}
\includegraphics[width=3.45in]{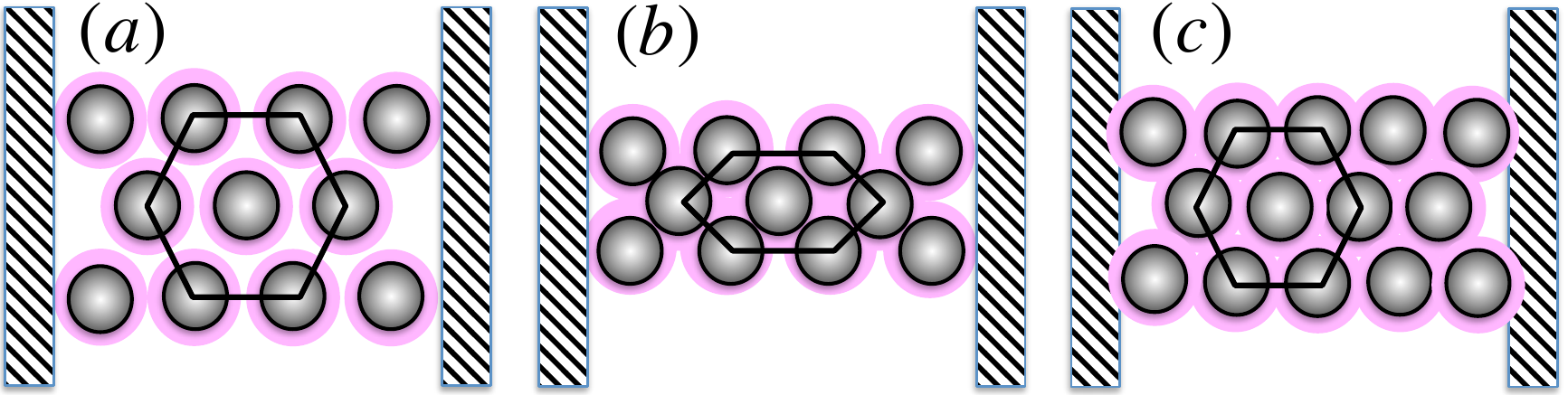}
\caption{\label{fig4} A dispersion with strong short-range order (a) can be compressed to a higher volume fraction by either a uniaxial deformation (b), or an isotropic deformation (c). The anisotropic structure in (b) can be reached through a simple affine deformation of the lattice whereas (c) requires the insertion of additional particles within each horizontal line of particles, and therefore a higher mobility. It is argued that the transformation from (a) to (c) is progressively blocked as the volume fraction increases in the soft colloidal glass.}
\end{figure}

In the case where the particle network resists changes in neighbours, we can discuss the magnitude of $\gamma$. The central idea is that the anisotropy arises from the uniaxial stress of the dispersion by the drag of the water past the particles. If the resulting strain was also purely uniaxial (as in an affine deformation of each particle's cage), such as shown in Fig. \ref{fig4}(a,b), then we would expect no further evolution of $q_x$, after the formation of cages at $\phi_c$.  This is not the case (Fig. \ref{fig2}b).  Further, the magnitude of the strain would be proportional to the change in volume fraction, since if $q_x$ remained constant, then $\phi/\phi_c = q_z/q_x$.  For our 8 nm silica in 5 mM NaCl, evolving from $\phi_c=0.35$ to $\phi=0.61$ would introduce an effective strain $\gamma=0.49$.  

\begin{figure}
\includegraphics[width=3.45in]{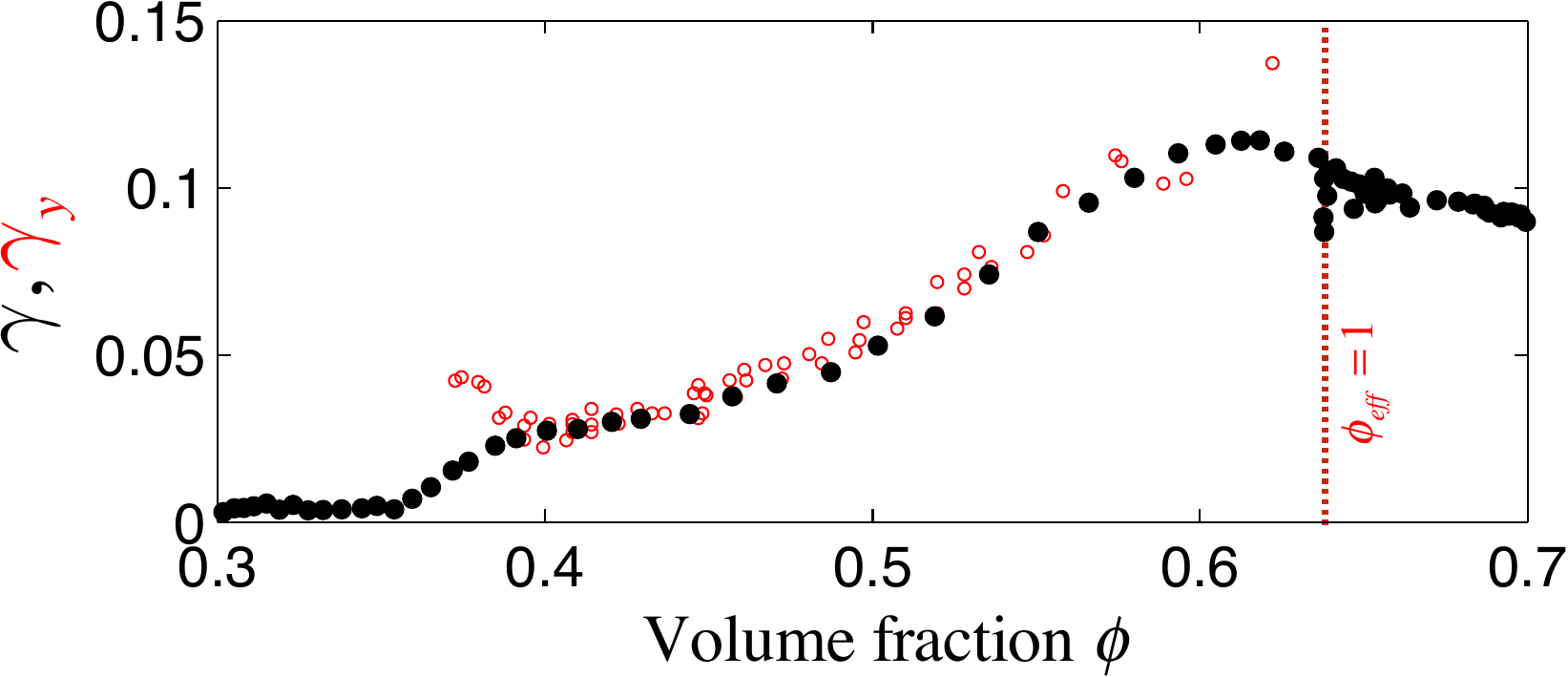}
\caption{\label{fig5} The effective strain $\gamma$ (black points) of the colloidal network increases as it dries, for 8 nm silica in 5 mM NaCl.  The material starts to behave as a colloidal glass at $\phi=0.35$, and accumulates strain until solidification at $\phi =0.61$.  This is in effect a graph of the largest deviatoric strain that the colloidal dispersion is able to retain, as a function of its volume fraction, i.e. the yield strain of the dispersion.  Superimposed, we show the measured yield strain (open circles) of sheared HIPR emulsions \cite{Mason1996}, with the $\phi$-axis rescaled such that the effective close packing of the emulsion $\phi_{eff}=1$ is mapped to the random close packing of hard spheres, $\phi=0.64$. }
\end{figure}

Clearly, despite the colloidal glass's resistance to shear, most of the uniaxial strain has actually been relieved.  As sketched in Fig. \ref{fig4}(c), relaxing this strain requires reordering the particles, and a change in the number of particles across the cell. This can be accomplished either by collective shear slip motions, or by local motions of individual particles.  Accordingly, as long as the applied strains are sufficiently large, we would expect the soft colloidal glass to evolve along some yield strain curve $\gamma = \gamma_y(\phi)$ during its compression. As such, in Fig. \ref{fig5} we replot the strain $\gamma$ against volume fraction, for 8 nm silica in 5 mM NaCl.    In the same figure, we also show the equivalent yield strain of sheared HIPR (high internal phase ratio) emulsions \cite{Mason1996}.  When concentrated, such emulsions are known to form soft cages around droplets that resist deformation \cite{Mason1996,Hebraud1997}. Interestingly, both the magnitude and shape of the yield strain curves of HIPR emulsions and our dispersions are quite comparable, if the maximum packing fraction $\phi_{eff}=1$ of the liquid emulsion droplets is mapped to $\phi=0.64$, the random close packing of spheres.  This may suggest that the yield strains here result from some general, geometric argument. 

Near the end of the colloidal glass region (for our 8 nm silica in 5 mM NaCl, from $\phi = 0.49$) the structure factor peak drops abruptly to half its original height, as shown in Fig. \ref{fig2}(d). At $\phi = 0.49$ the surface-surface distance of 8 nm particles is 1 nm, within the range of chemical interactions. Therefore the particles will begin to take positions that are dictated by reactions and non-central forces between the surfaces of the silica particles. This stage ends at the solidification front, when most particles have aggregated. At this point three things happen simultaneously: (a) the evolution of $q_x$ and $q_z$ nearly stops, indicating that the structure is no longer easily compressible; (b) the height $S_{max}$ of the structure factor peak goes to a much lower rate of change; and (c) the anisotropy $\gamma$ peaks, indicating that the affine deformation of the dispersion that created this anisotropy is no longer active. These observations can be explained by a large-scale aggregation of the particles into a comparatively incompressible aggregated network.

\begin{figure}
\includegraphics[width=3.45in]{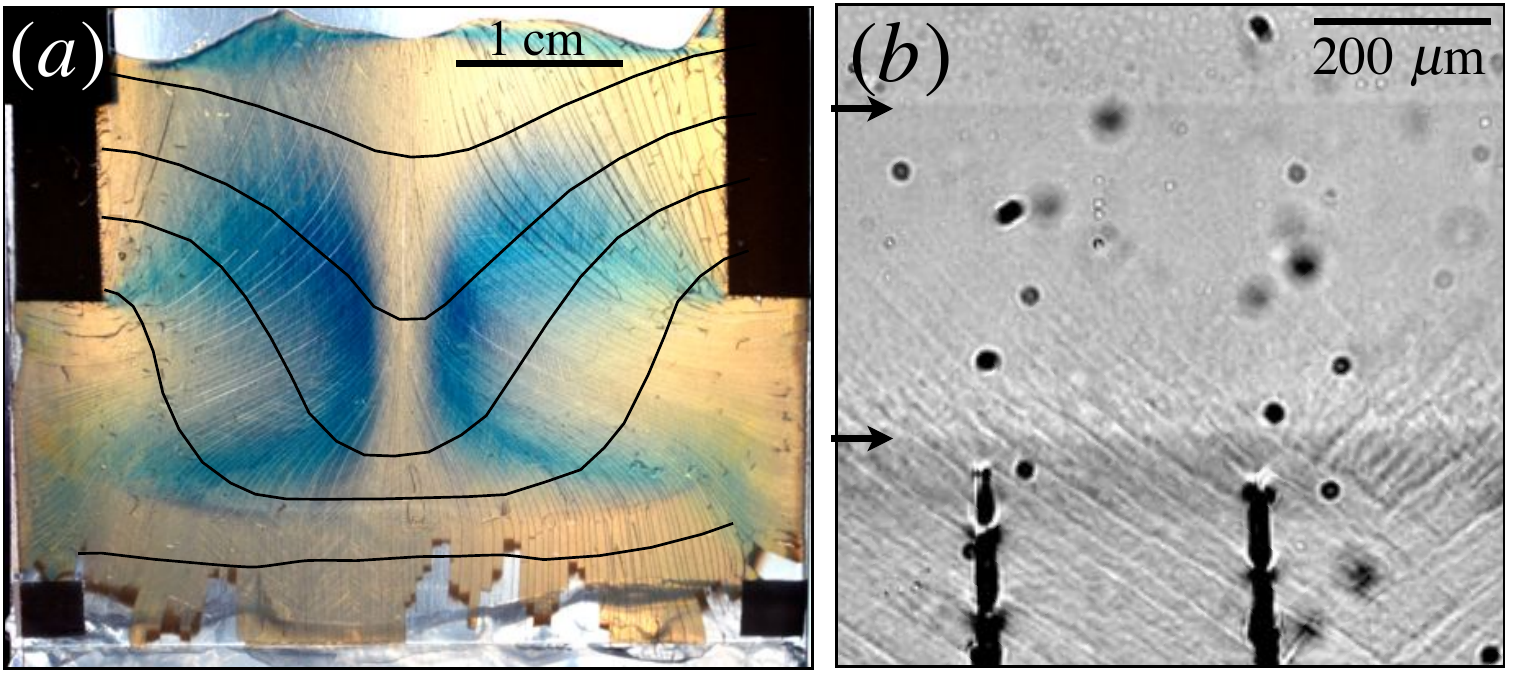}
\caption{\label{fig6} Consequences of structural anisotropy. (a) A dry cell of Levasil-30, between crossed polarisers, shows a pattern of birefringence that follows the direction of solidification.  Curves show the position of the solidification front every 100 minutes; drying occurred from the lower edge and sides of the cell, through the gaps in the (black) spacers. (b) Optical micrograph of a drying cell of 8 nm silica in 5 mM NaCl.  The caging and solidification fronts (arrows) are visible.  Shear bands are seen as lines $\sim\pm$45$^\circ$ to the solidification front, but precede it. }
\end{figure}

A remarkable feature of the liquid-glass-solid path to solidification here is the accumulation and transfer of strain to the solid, where it can influence the solid's properties. For example, optical birefringence is seen in dried colloidal droplets \cite{Inasawa2009,Yamaguchi2013}, and we have systematically seen birefringence in the directionally dried colloidal materials used here.  Such effects are expected when the packing arrangement, and hence refractive index, is anisotropic.  As shown in Fig. \ref{fig6}(a), the direction of birefringence can be controlled through the direction of solidification.

Yielding during this transition may also explain the appearance of periodic banding in drying dispersions. These have sometimes been called shear bands \cite{Hull1999,Berteloot2012} as their pattern looks similar to shear banding in metals, and form at $\sim \pm$45$^\circ$ to the direction of solidification.  As shown in Fig. \ref{fig6}(b), these bands actually appear in the transition region prior to solidification where they can act to transfer compression from the $z$-direction to the $x$-direction, through shear.  Although the connection is tentative, this is a potential driving force for such features, as this motion would act to relieve the accumulated $\gamma$.

After solidification some compression continues slowly in all the samples we observed.  The buildup of stress, and its release by cracks, in this elastic solid phase is well-studied (see e.g. \cite{Chiu1993a,Dufresne2003,Tirumkudulu2005,Routh2013}).  Other than a slight disturbance in the data of Fig. \ref{fig2} at $z = 1$ mm (which is correlated with the position of the drying front when the sample was installed in the beamline), the structural anisotropy relaxes slightly during this period.  This is likely due to cracks that open parallel to the $z$-axis during drying, and which compress the solid along the direction normal to their surface.  Whether there is any more direct interaction of the cracks with the structural anisotropy is perhaps an interesting future line of inquiry.

\section{Conclusion} We have shown that directional drying can induce structural anisotropy in colloidal dispersions. Anisotropy starts when the particles are caged by their mutual interactions.  The dispersion then acts as a yield-stress material, and accumulates additional strain in the direction of solidification, which is ultimately frozen into the structure of the aggregated solid.  If this is the case, it must be generic for all solid materials that are made through evaporation of dispersions that contain particles with reactive surfaces, which is the case of many coatings and ceramics. To our knowledge, such bulk structural anisotropy has never been reported for colloidal materials, although it is a robust feature of drying, and affects the optical and mechanical properties of the final solid.

\section{Acknowledgements}
We would like to thank ESRF and ILL for use of their facilities.


%

\end{document}